\input{psfig}

\documentstyle[prl,aps,twocolumn]{revtex}

\catcode`\@=11

\def\maketitle2{\par 
\begingroup
\let\cite\@bylinecite
\def\thefootnote{\fnsymbol{footnote}}%
\twocolumn[\@maketitle2\vskip2pc]%
\thispagestyle{plain}\@thanks
\endgroup
\def\thefootnote{\arabic{footnote}}%
\setcounter{footnote}{0}%
\let\maketitle2\relax \let\@maketitle2\relax
\let\@thanks\relax \let\@authoraddress\relax \let\@title\relax
\let\@date\relax \let\thanks\relax \let\@abstract\relax 
\let\@pacs\relax}

\def\abstract#1{\gdef\@abstract{{\par 
\bgroup
\ifdim\prevdepth=-1000pt \prevdepth0pt\fi
\hsize\columnwidth
\dimen0=-\prevdepth \advance\dimen0 by17.5pt \nointerlineskip
\small\vrule width 0pt height\dimen0 \relax}{~~}#1\egroup}}

\def\pacs#1{\gdef\@pacs{{\par 
\bgroup
\hsize\columnwidth \parindent0pt
\ifdim\prevdepth=-1000pt \prevdepth0pt\fi
\dimen0=-\prevdepth \advance\dimen0 by20pt\nointerlineskip
\egroup} PACS numbers:~#1}}

\def\@maketitle2{
\@preprint
\@title
\ifdim\prevdepth=-1000pt \prevdepth0pt\fi
\@authoraddress
\@date
\begin{list}{}{\leftmargin=0.10753\textwidth \rightmargin=\leftmargin
\itemsep=1pc\partopsep=-1pc}
\item\@abstract
\item\@pacs
\end{list}
}

\catcode`\@=12

\begin{document}

\title{\vspace*{-0.5cm}
\hspace*{\fill}{\normalsize LA-UR-97-4230} \\[1.5ex]
On the Equation of State of Nuclear Matter in
158A GeV Pb+Pb Collisions
}
\author{
B.R. Schlei${}^1$\thanks{E. Mail: schlei@t2.LANL.gov}{\ },
D. Strottman${}^1$\thanks{E. Mail: dds@LANL.gov}{\ }, and
N. Xu${}^2$\thanks{E. Mail: nxu@LBL.gov}{\ }
\\[1.5ex]
{\it ${}^1$Theoretical Division, Los Alamos National Laboratory, Los
Alamos, NM 87545}\\
{\it ${}^2$Nuclear Science Division, Lawrence Berkeley National Laboratory,
Berkeley,
CA 94720}
}
\date{October 20, 1997}

\abstract{
Within a hydrodynamical approach we investigate the sensitivity of single
inclusive momentum spectra of hadrons in 158A GeV Pb+Pb collisions to three
different equations of state of nuclear matter. Two of the equations of state
are based on lattice QCD results and include a phase transition to a quark-gluon 
plasma. The third equation of state has been extracted from the microscopic
transport code RQMD under the assumption of complete local thermalization.
All three equations of state provide reasonable fits to data taken by the NA44 and
NA49 Collaborations. The initial conditions before the evolution of the
fireballs and the space-time evolution pictures differ dramatically for the
three equations of state when the same freeze-out temperature is used in all
calculations. However, the softest of the equations of state results in 
transverse mass spectra that are too steep in the central rapidity region.
We conclude that the transverse particle momenta are determined by the
effective softness of the equation of state during the fireball expansion.}
\maketitle2
\narrowtext

A goal of studying ultrarelativistic heavy ion collisions is to
understand nuclear matter under extreme conditions such as high energy
density, high particle density, and whether a transition from a
quark-confined hadron state to the deconfined state (quark-gluon plasma,
QGP) takes place (see, e.g., ref.  \cite{qm96}). 
A highly excited state of matter and a
possible transition to a QGP will occur at the early stages of the
heavy-ion collisions; these differences will be reflected in different
equations of state for nuclear matter.  These in turn will affect the
maximum energy density and compression achieved in a heavy ion
reaction.  To investigate the nature of the hadronic equation of state
and possible phase transitions, it is very important to obtain
information about the conditions that exist at maximum compression,
the subsequent evolution of the system, and the final free hadrons
that can be measured in the detectors.

The equation of state, (EOS), is part of a hydrodynamic model, and the
connection between the EOS and experimental observables is relatively
direct.  In order to address issues rased above, we shall use the
simulation code HYLANDER-C ({\it cf.}, ref.  \cite{bernd15} and
refs. therein), with different equations of state and Cooper-Frye
freeze-out \cite{cooper}, to predict the single particle momentum 
distributions. The experimental spectra from the 158A GeV Pb+Pb collisions, 
measured by the NA44\cite{NA44xu,NA44baerden} and NA49 Collaborations
\cite{NA49jones}, are used as fit criteria for the hydrodynamical
(one-fluid type) calculations.

In our approach, virtually any type of EOS can be considered by solving the
relativistic Euler-equations\cite{euler}.  The coupled system of
partial differential equations necessary that describe the dynamics of
a relativistic fluid (with given initial conditions) contains an
equation of state which we write in the form 

\begin{equation}
P(\epsilon,n) \:=\: c^2(\epsilon,n)\,\epsilon\:.  \label{eq:press}
\end{equation}

In eq. (\ref{eq:press}), the quantities $P$, $\epsilon$, and $n$ are the
pressure, the energy density, and the baryon density, respectively.  The
proportionality constant, $c^2$, is in general a function of $\epsilon$ and
$n$.  In the following, we shall assume that in particular the $n$ dependence
is negligible for the energy regime under consideration.  From only the
knowledge of the speed of sound, $c_o^2(\epsilon)$, one can then calculate
$c^2(\epsilon)$ by evaluating the integral \cite{phdudo}
\begin{equation} 
c^2(\epsilon) \: =\: \displaystyle{\frac{1}{\epsilon}
\int_0^\epsilon}\: c_o^2(\epsilon^\prime)\:d\epsilon^\prime\:. \label{eq:c2}
\end{equation}

Assuming an adiabatic expansion, the temperature, $T(\epsilon)$, can be
calculated from \cite{phdudo}

\begin{equation}
T(\epsilon)\:=\:T_0\:\exp\left[
\displaystyle{\int_{\epsilon_0}^\epsilon}\:
\frac{c_o^2(\epsilon^\prime)\:d\epsilon^\prime}
{(1\:+\:c^2(\epsilon^\prime))\:\epsilon^\prime}
\right] \:,
\label{eq:temp}
\end{equation}

where $T_0=T(\epsilon_0)$ has to be specified for an arbitrary value of
$\epsilon_0$.

Three equations of state are discussed in this study. The first one,
EOS-I, which we use in the following exhibits a phase transition to a
quark-gluon plasma at a critical temperature $T_c$ = 200 $MeV$ with a
critical energy density $\epsilon_c$ = 3.2 $GeV/fm^3$ ({\it cf.},
ref. \cite{redlich}, and the refs. in \cite{bernd15}).  The second
equation of state, EOS-II, is also a lattice QCD-based EOS which has
recently become very popular in the field of relativistic heavy-ion
physics ({\it cf.}, ref. \cite{hung}).  This equation of state
includes a phase transition at $T_c$ = 160 $MeV$ with a critical
energy density $\epsilon_c$ $\approx$ 1.5 $GeV/fm^3$.  The third
equation of state, EOS-III, has been extracted from the microscopic
transport model RQMD \cite{rqmd21} under the assumption of complete
local thermalization \cite{sorge}, and does {\it not} include a
transition to a QGP.  In Fig. 1 the three equations of state are
plotted in many different representations.  For instance, in Fig. 1(d)
the plot of $c^2 = P(\epsilon)/\epsilon$ emphasizes the existence of a
minimum in $P/\epsilon$ at $\epsilon$ = $\epsilon_c$ = 3.2 $GeV/fm^3$
($\approx$ 1.5 $GeV/fm^3$) for EOS-I (EOS-II), referred to as {\it the
softest point} of the EOS.  It corresponds to the boundary between the
generalized mixed phase and the QGP \cite{hung}.  As seen in Fig. 1,
EOS-II yields a much softer equation of state than does EOS-I.

The parametrizations for the speed of sound which we have used here
are \cite{phdudo} 

\begin{eqnarray} c_o^2(\epsilon) &=&
[\alpha\:+\:\beta\:\tanh(\gamma \ln(\epsilon / \epsilon_0)+\delta)]
\nonumber\\ &&\times \left[1-\frac{(1-\xi)\Gamma^2}{\ln^2( \epsilon /
\epsilon_0 )+\Gamma^2} \right]\:, \label{eq:c02I} 
\end{eqnarray}

in the case of EOS-I;

\begin{eqnarray}
c_o^2(\epsilon) &=& \alpha\:+\:\beta\:\tanh[\gamma
\ln(\epsilon / \epsilon_0)]
\nonumber\\
&&-\:\delta\:\tanh[\xi\ln(\epsilon / \epsilon^\prime)]\:,
\label{eq:c02II}
\end{eqnarray}

in the case of EOS-II; and
\begin{equation}
c_o^2(\epsilon) \:=\:
\left\{
\begin{array}
{l@{\quad:\quad}l}
\alpha - \beta \left(
\epsilon / \epsilon_0 \right)^\delta
& \epsilon \leq \epsilon^\prime\\
0 & \epsilon > \epsilon^\prime
\end{array}
\right. \:,
\label{eq:c02III}
\end{equation}

in the case of EOS-III.  The parameters of eqs. (\ref{eq:c02I}) -
(\ref{eq:c02III}) and the choices for $T_0=T(\epsilon_0)$ are listed
in Table I.

In the following it is assumed that due to an experimental uncertainty
for the centrality of the collision, only 90$\%$ of the total
available energy and the total baryon number have been observed.  It
is then possible to find initial distributions for the three equations
of state, such that one can reproduce the single inclusive momentum
spectra of 158 $AGeV$ Pb+Pb collisions ({\it cf.}, Fig. 3).

Fig. 2 shows the initial distributions for the energy density,
$\epsilon(z)$, the baryon density, $n_B(z)$, and the fluid rapidity,
$y_F(z)$, plotted against the longitudinal coordinate $z$.  We use here
for the initial distributions the initial condition scenarios which
have been extensively described in refs. \cite{bernd8} and \cite{jan}.
In particular, it is assumed that an initial transverse fluid velocity
field is absent, and the initial longitudinal distributions for energy
density, $\epsilon$, and baryon density, $n_B$, are smeared out with a
Woods-Saxon parametrization in the transverse direction, $r_\perp$
({\it cf.}, ref.  \cite{bernd3}). The choices for the initial
parameters, which in each case provide the best fit results for the
hadronic single inclusive momentum spectra of the two considered
experiments, are listed in Table II.

The maximum initial energy density is $\epsilon_\Delta$ = 15.3 (6.55
(8.66)) $GeV/fm^3$ , and the maximum initial baryon density is
$n_B^{max}$ = 3.93 (1.24 (1.92)) $fm^{-3}$, for EOS-I (EOS-II
(EOS-III)), respectively; 71\% (30\% (36\%)) of the baryonic matter is
initially located in the central fireball region.  Hence, EOS-I predicts 
a much larger stopping than EOS-III does, and EOS-III predicts a much 
larger stopping than EOS-II does.

Fig. 3 shows results of the hydrodynamical calculations compared to
the single inclusive momentum spectra of the 158 $AGeV$ Pb+Pb
collisions which have been measured by the NA44
\cite{NA44xu},\cite{NA44baerden} and NA49 Collaborations
\cite{NA49jones}.  All single inclusive momentum spectra have been
evaluated in the nucleus-nucleus center of mass system ($y_{cm}$ =
2.91), and for all calculations the freeze-out temperature $T_f$ = 139
$MeV$ has been used.  The energy density at freeze-out is $\epsilon_f$
= 0.292 (0.126 (0.130)) $GeV/fm^3$ for EOS-I (EOS-II (EOS-III)).  In
particular, the freeze-out energy density, $\epsilon_f$, is a factor
$\sim 2.2$ larger in the calculation using EOS-I compared to the other
two calculations.  Inspecting Fig. 4, we found that the total lifetime
of the fireball is reduced by approximately this factor in the
calculation using EOS-I.  Fig. 4 also shows that EOS-II and EOS-III
yield fireballs of similar lifetimes ({\it cf.} also Table II).  These
particular features show up in Bose-Einstein correlation (BEC)
calculations ({\it cf.} ref. \cite{bernd15}); more on BEC will be
published elsewhere \cite{future}. The relatively strong effect in
two-particle correlation functions was also observed in a recent
report \cite{gyulassy}.

All three calculations yield resonable agreement with both experiments.
However, the calculations for the very soft EOS-II yield transverse
mass spectra that are too steep in the central rapidity region.  The
magnitudes of the slopes in the transverse mass spectra have their origin 
in the transverse velocity fields at freeze-out.  For EOS-II its
corresponding transverse velocity field is the weakest ({\it cf.}
Table II) of the three EOS.  In fact, it is about 30\% smaller than
the transverse velocity fields for EOS-I and EOS-III. The final
velocity is proportional to the time integral

\begin{equation}
\int_{t_i}^{t_f} \frac{P}{\epsilon}(t^\prime)\:d t^\prime \:,
\label{eq:mean}
\end{equation}

where $P/\epsilon$ is proportional to the fluid acceleration
\cite{shuryak1,shuryak2}. In Fig. 1(d) we exhibit for each EOS we used
the starting values of $P/\epsilon$ with respect to the initial
maximum energy density $\epsilon_\Delta$ at transverse position
$r_\perp = 0$, and the final values of $P/\epsilon$ at breakup energy
densities, $\epsilon_f$.  For the given total fireball lifetime,
$t_{max} \equiv t_f - t_i$, the effective $P/\epsilon$ of EOS-II is
obviously too small, whereas $P/\epsilon$ of EOS-I and EOS-III appear
to have the correct effective softness.  We stress that the total
lifetime, $t_{max}$, in the case of EOS-I is reduced by a factor $\sim
2.1$ compared to $t_{max}$ of EOS-III, but the ratio $P/\epsilon$ of
EOS-I is effectively increased for the considered expansion period.

In summary, we have tested three types of equation of state for 158A
GeV Pb+Pb collisions. We found that the measured single inclusive
momentum spectra are sensitive to both the initial conditions and to
the equation of state.  EOS-I and EOS-III both provide reasonable fits
to the data.  For the EOS-II a softest point is assumed and its
corresponding velocity field is the weakest.  As a result, the
transverse mass distributions calculated from this equation of state
are found to be too steep in the central rapidity region as compared
to the experimental observations. Obviously, the transverse particle
momenta are sensitive to the effective softness of the equation of
state during the fireball expansion.  

As mentioned earlier, our conclusions are based on one-fluid
hydrodynamic model calculations. Furthermore, the freeze-out
temperature is fixed at 139 MeV for all three cases under study. As
long as we limit our comparison of different scenarios within the same
framework, our conclusions are valid. Still, several crucial questions
like where and when the collective velocity field \cite{larry} is
developed during the collision, and whether it develops in the QGP phase 
or in the hadronic phase are not answered yet. In addition, the exact 
condition for the system evolving from interacting hadrons to 
freeze-streaming ones is not known. To shed light in this direction, 
an even more sophisticated analysis is needed. We believe that the combined
analysis of single inclusive momentum spectra and Bose-Einstein
correlations will help to further constrain\cite{ferenc} the limits of
possible candidates for the equation of state of nuclear matter in
relativistic heavy-ion collisions.

We thank Dr. H. Sorge for providing us all the neccessary information
for the construction of the RQMD EOS.  We are grateful for many
enlightening discussions with Drs. M. Prakash, C. Redlich, D. Rischke,
E.V. Shuryak, and H. Sorge.  Special thanks goes to M. Toy for
providing us data files of the preliminary NA49 data.  This work has
been supported by the U.S. Department of Energy.

\vspace*{-0.5cm}
\begin{figure}
\begin{center}\mbox{ }
\hspace*{-0.35cm}
{\psfig{file=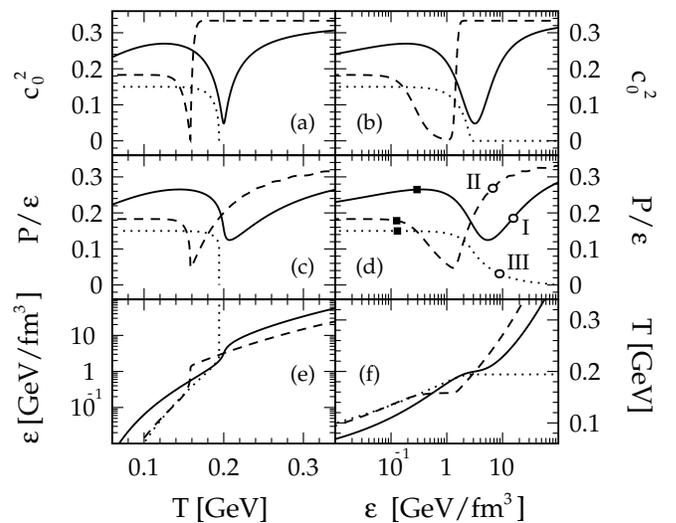,width=9.1cm}}\end{center}
\vspace*{0.0cm}
\caption{
Energy density, $\epsilon$, ratio of pressure and energy density,
$P/\epsilon$, speed of sound, $c_0^2$, and temperature, $T$, as functions of
$T$ and/or $\epsilon$, for the equations of state EOS-I (solid lines),
EOS-II (dashed lines), and EOS-III (dotted lines), respectively (see text).
The open circles in plot (d) correspond for each EOS to the starting values
of
$P/\epsilon$ with respect to the achieved initial maximum energy density
$\epsilon_\Delta$ at transverse position $r_\perp = 0$, whereas the filled
squares represent the final values of $P/\epsilon$ at breakup energy densities,
$\epsilon_f$.  }
\label{fg:fig1}
\end{figure}

\cleardoublepage

\vspace*{-1.5cm}
\begin{figure}
\begin{center}\mbox{ }
\hspace*{-0.3cm}
{\psfig{file=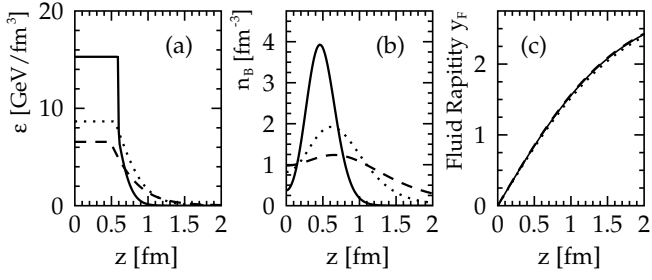,width=9.0cm}}
\end{center}
\vspace*{0.0cm}
\caption{
Initial distributions of (a) the energy density, $\epsilon$, (b) the baryon
density, $n_B$, and (c) the fluid rapidity, $y_F$, plotted against the
longitudinal coordinate $z$ at transverse position $r_\perp = 0$.  The
solid (dashed (dotted)) lines indicate the initial distributions
for the calculation using EOS-I (EOS-II (EOS-III)).  }
\label{fg:fig2}
\end{figure}

\vspace*{0.0cm}
\begin{figure}
\begin{center}\mbox{ }
\hspace*{-0.5cm}
{\psfig{file=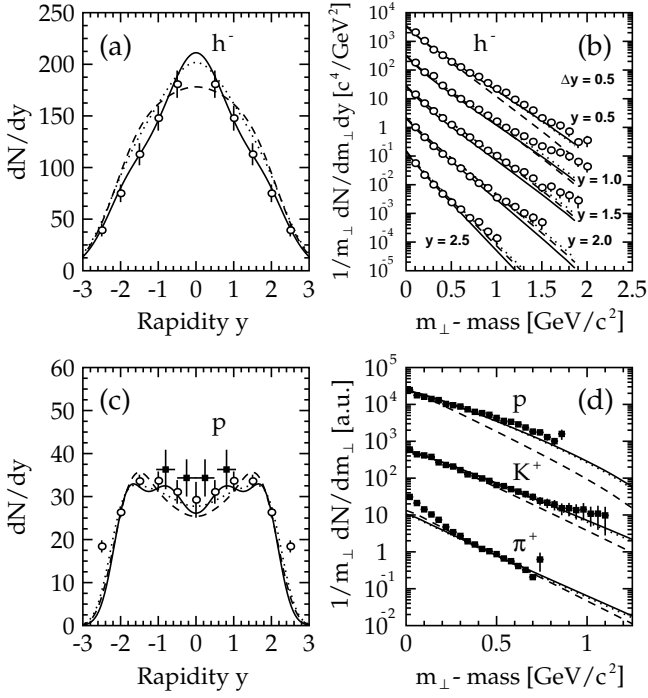,width=9.2cm}}
\end{center}
\vspace*{0.0cm}
\caption{
(a) Rapidity spectra and (b) transverse mass spectra,
$1/m_\perp dN/dm_\perp dy$, of negative hadrons, $h^-$, (c) rapidity spectra
of protons (without contributions from $\Lambda^0$ decay) and (d) transverse
mass spectra, $1/m_\perp dN/dm_\perp$, of protons (including contributions
from
$\Lambda^0$ decay), $p$, positive kaons, $K^+$, and positive pions, $\pi^+$,
respectively.  The solid (dashed (dotted)) lines indicate the results of the
calculations when using equation of state EOS-I (EOS-II (EOS-III)).  The open
circles represent preliminary data taken by the NA49 Collaboration, whereas
the filled squares represent final data taken by the NA44 Collaboration.  }
\label{fg:fig3}
\end{figure}

\newpage

\vspace*{-2.0cm}
\begin{figure}
\begin{center}\mbox{ }
\hspace*{-0.3cm}
{\psfig{file=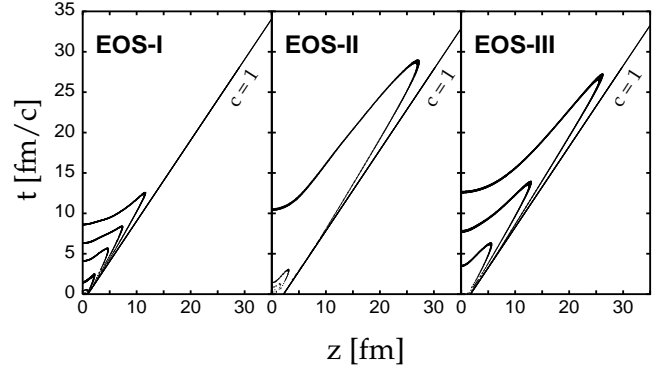,width=9.0cm}}
\end{center}
\vspace*{0.0cm}
\caption{
Isothermes for the relativistic fluids governed by EOS-I, EOS-II and EOS-III
at $r_\perp = 0$, respectively.  The outer lines correspond to a temperature,
$T$ = 139 $MeV$, and each successively smaller curve represents a reduction
in temperature by $\Delta T$ = 20 $MeV$.  The lines $c=1$ represent the light
cone.}
\label{fg:fig4}
\end{figure}

\vspace*{-0.5cm}
\begin{table}
\caption{EOS Parameters.}
\begin{center}
\begin{tabular}{l c c c}
 & EOS-I & EOS-II & EOS-III \\
\hline
$\alpha$             & 4/15 & 0.2582 & 0.15 \\
$\beta$              & 1/15 & 1/6  & 0.02 \\
$\gamma$             & 0.24 & 10.0  & $-$  \\
$\delta$             & 1.05 & 0.0915 & 2.00 \\
$\xi$               & 0.15 & 1.50  & $-$  \\
$\Gamma^2$            & 0.73 & $-$  & $-$  \\
$\epsilon^\prime$ $[GeV/fm^3]$  & $-$  & 0.27  & 2.80 \\
$\epsilon_0$ $[GeV/fm^3]$     & 3.20 & 1.45  & 1.00 \\
$T_0$ $[MeV]$           & 200  & 160  & 180  \\
\end{tabular}
\end{center}
\end{table}

\vspace*{-1.0cm}
\begin{table}
\caption{Properties of the fireballs.}
\begin{center}
\begin{tabular}{p{4.5cm} c c c}
 & EOS-I & EOS-II & EOS-III \\
\hline
\multicolumn{4}{c}{Initial parameters}\\
Rel.  fraction of thermal energy in the central fireball, $K_L$
& 0.55 & 0.20 & 0.28 \\
Longitudinal extension of the central fireball, $\Delta$ $[fm]$
& 1.20 & 1.00 & 1.10 \\
Rapidity at the edge of the central fireball, $y_\Delta$
& 1.00 & 0.85 & 0.90 \\
Rapidty at maximum of initial baryon $y$ distribution, $y_m$
& 0.80 & 1.50 & 1.10 \\
Width of initial baryon $y$ distribution, $\sigma$
& 0.32 & 1.00 & 0.60 \\
\multicolumn{4}{c}{Output}\\
Max.  initial energy density, $\epsilon_\Delta$ $[GeV/fm^3]$
& 15.3 & 6.55 & 8.66 \\
Max.  initial baryon density, $n_B^{max}$ $[fm^{-3}]$
& 3.93 & 1.24 & 1.92 \\
Rel.  fraction of baryons in central fireball, $f_{n_B}^\Delta$
& 0.71 & 0.30 & 0.36 \\
Freeze-out energy density, $\epsilon_f$ $[GeV/fm^3]$
& 0.292 & 0.126 & 0.130 \\
Max.  transverse velocity at freeze-out, $v_\perp^{max}$ $[c]$
& 0.46 & 0.30 & 0.42 \\
Lifetime of fireball, $t_{max}$ $[fm/c]$
& 13.1 & 29.3 & 27.3 \\
Lifetime of QGP, $t_{QGP}$ $[fm/c]$
& 2.4  & 3.0  & $-$ \\
\end{tabular}
\end{center}
\vspace*{-0.5cm}
\end{table}

\end{document}